\documentclass[pra,twocolumn,superscriptaddress]{revtex4-2} 
\usepackage{mathrsfs}
\usepackage{amsfonts}
\usepackage{amsmath}
\usepackage{amssymb}
\usepackage{graphicx,subfigure}
\usepackage{bm}
\usepackage{color}
\usepackage[normalem]{ulem}
\usepackage{dcolumn} 
\usepackage[table]{xcolor}
\usepackage{float}
\usepackage{tabularx}
\usepackage{hyperref}
\newcommand{\ket}[1]{|#1\rangle}
\newcommand{\bra}[1]{\langle #1|}

\begin{document}

\title{Separation of quantum time evolution into holonomic and dynamical parts} 

\author{Adam Fredriksson}
\email{Adam.Fredriksson@uni-siegen.de}
\affiliation{Department of Physics and Astronomy, Uppsala University, 
Box 524, SE-751 20 Uppsala, Sweden}
\affiliation{Naturwissenschaftlich-Technische Fakult\"at, Universit\"at Siegen, 
Walter-Flex-Stra{\ss}e 3, D-57068 Siegen, Germany}
\author{Erik Sj\"oqvist}
\email{erik.sjoqvist@physics.uu.se}
\affiliation{Department of Physics and Astronomy, Uppsala University, 
Box 524, SE-751 20 Uppsala, Sweden}

\date{\today}

\begin{abstract}
The issue of separating Schr\"odinger-type quantum time evolution into a product of holonomic 
and dynamical parts in the non-adiabatic non-Abelian case is examined. 
We identify all special cases in which this kind of separation is possible, and we prove that 
separability is a gauge invariant property of quantum time evolution.
The general analysis is implemented in a three-level 
system with $\Lambda$-type coupling structures. The typical case where the holonomic 
and dynamical parts do not separate is illustrated by means of two non-commuting $\Lambda$-type 
Hamiltonians. 
\end{abstract}

\maketitle

\section{Introduction}
Experimental tests of various forms of geometric phase \cite{berry84,aharonov87,samuel88} 
and quantum holonomy \cite{wilczek84,anandan88,kult06} rely on the ability to separate 
it from the dynamical part of the quantum time evolution 
\cite{suter88,zwanziger90,wagh97,jones99,toyoda13,leroux18}. While such a product form 
always exists in the Abelian geometric phase cases \cite{berry84,aharonov87} and in the 
case of adiabatic non-Abelian quantum holonomy \cite{wilczek84}, the holonomic and 
dynamical parts do not generally separate for non-adiabatic time evolution of subspaces 
of Hilbert space \cite{anandan88}; a setting that has gained considerable attention in the 
context of holonomic quantum computation 
\cite{sjoqvist12,abdumalikov13,feng13,arroyo14,zu14,xu15,sjoqvist16, zhou17,herterich16,xu18,zhang23}. 
Motivated by the recent analyses in Refs.~\cite{zhao23,yu23,fredriksson26,yu26}, we undertake 
a detailed examination of the issue of separation for non-adiabatic Schr\"odinger-type quantum 
time evolution of subspaces. 

The theory of quantum holonomy in the non-adiabatic non-Abelian case was first developed 
by Anandan \cite{anandan88}. Specifically, it was established that the Schr\"odinger evolution 
of a subspace of Hilbert space is governed by a time-ordered exponential of a sum of a matrix-valued 
vector potential and a matrix representation of the Hamiltonian, generating the holonomic and 
dynamical parts of the time evolution, respectively. The generators are typically non-commuting 
matrices, which implies that the quantum time evolution cannot, in general, be separated into 
a product of a holonomic and a dynamical part, thereby limiting the experimental accessibility 
of quantum holonomy.

Although separation into holonomic and dynamical parts typically does not occur, it remains 
pertinent to identify the special cases in which it does occur for non-adiabatic time evolution 
of subspaces. These cases are of particular interest for experimental tests of quantum holonomies 
in non-adiabatic evolution and are relevant for holonomic quantum computation. To this end, 
we revisit the theory of Schr\"odinger-type quantum time evolution of subspaces
to analyze the issue of separation systematically.

The paper is organized as follows. We commence in Sec.~\ref{sec:II} by presenting an 
operator-based approach to the Schr\"odinger evolution of subspaces and we show how 
it underlies the theory developed in Ref.~\cite{anandan88}. Notably, the analysis is carried 
out for non-cyclic time evolution throughout. We identify all cases in which separation of 
quantum time evolution into holonomic and dynamical parts does occur. We further prove 
that the separability is a gauge invariant property of quantum time evolution. In Sec.~\ref{sec:IV}, 
the general analysis is illustrated by a series of examples, all based on the three-level 
$\Lambda$ configuration. Finally, a brief summary of the paper and an 
outlook delineating potential directions for further study is provided in Sec.~\ref{sec:V}.

\section{Schr\"odinger evolution of subspaces}
\label{sec:II}
\subsection{The restricted Schr\"odinger equation}
Consider a closed quantum system associated with an $N$ dimensional ($N$ finite) 
Hilbert space $\mathscr{H} = \mathscr{V}_{M} (t) \oplus \mathscr{V}_{N-M}(t)$ decomposed 
into two orthogonal subspaces $\mathscr{V}_{M}(t)$ and $\mathscr{V}_{N-M}(t)$ of fixed 
dimensions $M$ and $N-M$, respectively. We assume that $C\!:t\in[0,\tau] \mapsto \mathscr{V}_{M} (t)$ 
is a continuous curve in the Grassmannian $\mathscr{G} (N;M)$, i.e., the space of $M$ dimensional 
subspaces of $\mathscr{H}$, and introduce an orthonormal $M$-frame 
$\mathcal{S}(t) \equiv \{\ket{\psi_{j}(t)}\}_{j=1}^{M}$ spanning $\mathscr{V}_{M}(t)$ at 
every instant $t$, where each $\ket{\psi_{j}(t)}$ is a solution of the time-dependent 
Schr\"odinger equation (we put $\hbar = 1$): 
\begin{eqnarray}
\ket{\dot{\psi}_{j}(t)} = -iH(t)\ket{\psi_{j}(t)}, \quad \ j=1,\ldots, M
\label{eq:se}
\end{eqnarray}
with $H(t)$ denoting the Hamiltonian operator of the system. The time evolution 
operator $U(t,0)\!:\mathscr{V}_{M}(0)\to\mathscr{V}_{M}(t)$ may be written as
\begin{eqnarray}
U(t,0) = \ket{\psi_{j}(t)}\bra{\psi_{j}(0)} . 
\label{eq:ytu}
\end{eqnarray}
Here and onwards, unless otherwise stated, we use Einstein's summation convention, according 
to which repeated indices are 
implicitly summed. The operator $U(t,0)$ is a partial isometry, i.e., $U^{\dagger}(t,0)U(t,0) = 
\ket{\psi_{j}(0)}\bra{\psi_{j}(0)} \equiv P_{M}(0)$ and $U(t,0)U^{\dagger}(t,0) = 
\ket{\psi_{j}(t)}\bra{\psi_{j}(t)}\equiv P_{M}(t)$ are projectors on the subspaces 
$\mathscr{V}_{M}(0)$ and $\mathscr{V}_{M}(t)$, respectively. The special case where 
$P_{M}(\tau) = P_{M}(0)$ defines cyclic evolution, i.e., when $\mathscr{V}_{M} (\tau)$ 
and $\mathscr{V}_{M} (0)$ coincide.
 
We proceed by emphasizing the general fact that all quantum holonomy experiments start 
from a physical setup that defines a Hamiltonian $H(t)$ and by preparing the initial projector 
$P_{M}(0)$ on $\mathscr{V}_{M}(0)$. The Schr\"odinger equation for the projector
\begin{eqnarray}
\dot{P}_{M} (t) = i [P_{M}(t), H(t)]
\label{eq:vonNeumann}
\end{eqnarray}
generates a continuous curve $\tilde{C}\!: t \in [0,\tau] \mapsto P_{M} (t)$ in the space of 
rank-$M$ projectors, being isomorphic to the curve $C$ in $\mathscr{G} (N;M)$ introduced 
above. Thus, the experimental setup determines $H(t)$ and $P_{M}(t)$. 

The essence of the preceding observation is captured by the differential equation 
\cite{leone19,remark1} 
\begin{eqnarray}
\dot{U}(t,0) = [\dot{P}_{M}(t) - iP_{M}(t)H(t)P_{M}(t)]U(t,0) , 
\label{eq:modyte}
\end{eqnarray}
where $P_{M}(t)H(t)P_{M}(t)$ is the Hamiltonian operator restricted to the instantaneous 
subspace $\mathscr{V}_{M}(t)$. We will refer to Eq.~\eqref{eq:modyte} as the restricted 
Schr\"odinger equation  (RSE).

\subsection{Derivation of the time evolution matrix}
We establish that Anandan's theory \cite{anandan88} follows from RSE. This requires the introduction 
of an orthonormal $M$-frame $\mathcal{L}(t)\equiv\{\ket{\varphi_{j}(t)}\}_{j=1}^{M}$ 
consisting of vectors that span $\mathscr{V}_{M}(t)$ at every instant $t$, but generally do not 
satisfy Eq.~\eqref{eq:se}, and obey the `in-phase' condition \cite{remark2}
\begin{eqnarray}
\mathbb{O}(0,\tau) > 0,
\label{eq:inphase}
\end{eqnarray}
where $\mathbb{O}(0,\tau)$ is the $M\times M$ overlap matrix with matrix elements 
$\mathbb{O}_{jk}(0,\tau) = \langle \varphi_j (0) \ket{\varphi_k (\tau)}$. For cyclic evolution 
$\mathscr{V}_{M}(\tau) = \mathscr{V}_{M}(0)$, Eq.~\eqref{eq:inphase} reduces to the standard 
condition $\ket{\varphi_{j}(\tau)} = \ket{\varphi_{j}(0)}$, in case of which 
$\mathbb{O}(0,\tau)=\mathbb{I}_{M} >0$, $\mathbb{I}_{M}$ being the $M\times M$ 
identity matrix. In the language of fiber bundles, $\mathcal{L}(t)$ corresponds to a local 
section \cite{bohm91} that serves as a reference used for a gauge covariant description 
of the change $\mathcal{S} (0) \mapsto \mathcal{S} (\tau)$ generated by the Hamiltonian $H(t)$.

The two $M$-frames $\mathcal{S} (t)$ and $\mathcal{L} (t)$ spanning $\mathscr{V}_{M} (t)$ 
are related by the operator $W(t)\equiv\ket{\psi_{j}(t)} \bra{\varphi_{j}(t)}$, which is a partial 
isometry as $W^{\dagger} (t) W(t) = W(t) W^{\dagger} (t) = P_{M}(t)$. In the following, we 
assume that $\ket{\varphi_{j}(0)} = \ket{\psi_{j}(0)}$, corresponding to the initial condition 
$W(0) = P_{M}(0)$. We may write
\begin{eqnarray}
\ket{\psi_{k}(t)} = 
\ket{\varphi_{j}(t)} \mathbb{W}_{jk}(t) ,
\label{eq:FramesRel}
\end{eqnarray}
where $\mathbb{W}_{jk}(t)\equiv\bra{\varphi_{j}(t)}W(t)\ket{\varphi_{k}(t)} = 
\langle\varphi_{j}(t)\ket{\psi_{k}(t)}$ are matrix elements of the unitary $M\times M$ 
transformation matrix $\mathbb{W}(t)$, which connects $\mathcal{L}(t)$ and $\mathcal{S}(t)$.
Equations \eqref{eq:ytu} and \eqref{eq:FramesRel}, together with 
$\mathbb{W}_{jk} (0) = \delta_{jk}$, yield
\begin{eqnarray}
U(t,0) & = & P_{M}(t) U(t,0) 
\nonumber \\ 
 & = & P_{M}(t) \ket{\varphi_j (t)} 
\bra{\varphi_k (0)} \mathbb{W}_{jk} (t) . 
\label{eq:UandW}
\end{eqnarray}
By inserting this expression into RSE, we find
\begin{eqnarray}
& & P_{M}(t) \ket{\dot{\varphi}_j (t)} \bra{\varphi_k (0)} \mathbb{W}_{jk} (t) + \ket{\varphi_j (t)} 
\bra{\varphi_k (0)} \dot{\mathbb{W}}_{jk} (t) 
\nonumber \\ 
 & = & -iP_{M}(t) H(t) \ket{\varphi_{j}(t)}\bra{\varphi_{k}(0)}\mathbb{W}_{jk}(t), 
\end{eqnarray}
where the term $\dot{P}_{M}(t)U(t,0)$ has been canceled on both sides. Further, upon 
substitution of $P_{M}(t) = \ket{\varphi_l (t)} 
\bra{\varphi_l (t)}$ and relabeling in the second term on the left-hand side, we have    
\begin{eqnarray}
 & & \ket{\varphi_l (t)} \bra{\varphi_k (0)}
\langle \varphi_l (t) \ket{\dot{\varphi}_j (t)} \mathbb{W}_{jk} (t) 
\nonumber \\ 
 & & + \ket{\varphi_l (t)} \bra{\varphi_k (0)} \dot{\mathbb{W}}_{lk} (t) 
\nonumber \\ 
 & = & -i \ket{\varphi_l (t)} \bra{\varphi_k (0)} \bra{\varphi_l (t)} H(t) \ket{\varphi_j (t)} 
 \mathbb{W}_{jk} (t) .  
\end{eqnarray}
By using the linear independence of the operators $\ket{\varphi_l (t)} \bra{\varphi_k (0)}$ 
and that $\langle \varphi_l (t) \ket{\dot{\varphi}_j (t)} = - \langle \dot{\varphi}_l (t) \ket{\varphi_j (t)}$, 
as well as rearranging terms, we obtain the Anandan equation (AE):
\begin{eqnarray}
\dot{\mathbb{W}}(t) = 
[\mathbb{A}(t) + \mathbb{K}(t)]\mathbb{W}(t).
\label{eq:ae}
\end{eqnarray}
Here, $\mathbb{A} (t)$ and $\mathbb{K}(t)$ are anti-Hermitian $M\times M$ matrices with 
matrix elements
\begin{eqnarray} 
\mathbb{A}_{jk}(t) & \equiv & \langle\dot{\varphi}_{j}(t)\ket{\varphi_{k}(t)}, 
\nonumber \\ 
\mathbb{K}_{jk} (t) & \equiv & -i\bra{\varphi_{j}(t)} H(t) \ket{\varphi_{k}(t)} . 
\label{eq:aeaf}
\end{eqnarray}
The formal solution of AE is given by a time-ordered exponential:   
\begin{eqnarray}
\mathbb{W}(\tau) = \mathcal{T}e^{\int_{0}^{\tau} [\mathbb{A}(t) + \mathbb{K}(t)] dt}  ,
\label{eq:ae_solution}
\end{eqnarray}
which typically does not separate into a product of holonomic $\mathcal{T} 
e^{\int_{0}^{\tau} \mathbb{A}(t) dt}$ and dynamical $\mathcal{T} e^{\int_{0}^{\tau} \mathbb{K}(t) dt}$ 
parts \cite{anandan88,zhao23,bohm03,duzzioni08,ericsson13}, 
as $\mathbb{A} (t)$ and $\mathbb{K} (t')$ generally 
do not commute for arbitrary $t,t' \in [0,\tau]$. Physically, the difference between 
$\mathcal{T} e^{\int_{0}^{\tau} \mathbb{A}(t) dt}$ and $\mathcal{T} e^{\int_{0}^{\tau} \mathbb{K}(t) dt}$ 
is that whereas the former is a time non-local path-dependent quantity \cite{remark3}, the latter is 
time local, i.e., it is given by the time integral of the Hamiltonian evaluated at time $t$ projected 
onto the subspace at the same time $t$ \cite{mukunda93}.  

The use of an $M$-frame $\mathcal{L}(t)$ satisfying the `in-phase' condition in Eq.~\eqref{eq:inphase} 
ensures that the time evolution must not be cyclic. To make this point precise, we note that 
\begin{eqnarray}
\mathbb{U}_{jk} (\tau,0)  
 & \equiv &  \bra{\psi_j (0)} U(\tau,0) 
 \ket{\psi_k (0)} = \langle \psi_j (0) \ket{\psi_k (\tau)} 
\nonumber \\ 
 & = & \langle \varphi_j (0) \ket{\varphi_l (\tau)} 
\mathbb{W}_{lk} (\tau) 
\nonumber \\ 
 & = & \mathbb{O}_{jl} (0,\tau) 
\mathbb{W}_{lk} (\tau), 
\label{eq:noncyclicmodAE}
\end{eqnarray}
where $\mathbb{U} (\tau,0)$ is the matrix representation of the time evolution operator 
$U(\tau,0)$ with respect to $\mathcal{S}(0)$ \cite{remark4}. The matrices $\mathbb{O} (0,\tau)$ 
and $\mathbb{W} (\tau)$ are the positive and unitary parts, respectively, of $\mathbb{U} (\tau,0)$. 
Thus, $\mathbb{U} (\tau,0)$ is unitary if and only if $\mathbb{O} (0,\tau) = \mathbb{I}_{M}$, which, 
as noted above, corresponds to cyclic evolution. By combining Eqs.~\eqref{eq:ae_solution} and 
\eqref{eq:noncyclicmodAE}, we obtain the time evolution matrix that governs the Schr\"odinger 
evolution of the subspace $\mathscr{V}_{M}(t)$:
\begin{eqnarray}
\mathbb{U}(\tau,0) = \mathbb{O}(0,\tau) \mathcal{T} e^{\int_{0}^{\tau} 
[\mathbb{A}(t) + \mathbb{K}(t)] dt},
\label{eq:genU}
\end{eqnarray}
which reduces to Anandan's expression \cite{anandan88} for cyclic evolution. Equation 
\eqref{eq:genU} shows that the issue of separation applies to any quantum evolution, 
irrespective of whether it is cyclic or non-cyclic. 

\subsection{The issue of separation}

\label{sec:IIC}
Although the preceding analysis establishes 
that quantum time evolution does not separate into a product of a holonomic and dynamical part in general, 
it remains pertinent to determine the special cases in which such a separation does 
exist for non-adiabatic time evolution of subspaces of dimension $M \geq 2$. The 
requirement for the solution of AE in Eq.~\eqref{eq:ae_solution} to separate is that
\begin{eqnarray}
[\mathbb{A} (t),\mathbb{K} (t')] = 0_M, \quad \forall t,t' \in [0,\tau] , 
\label{eq:AKcomm}
\end{eqnarray}
$0_M$ being the $M \times M$ zero matrix. 
This is a highly restrictive condition in the sense that it will typically not be satisfied for a 
randomly chosen Hamiltonian and subspace. To make this point precise, we identify the only 
three distinct special cases for the  requirement in Eq.~\eqref{eq:AKcomm} to hold: (i) when 
$\mathscr{V}_{M} (t) = 
\mathscr{V}_{M} (0)$, $\forall t \in [0,\tau]$, in which case just the dynamical part can be 
non-trivial, as the vectors defining the $M$-frame $\mathcal{L}(t)$ can be chosen 
time-independent; (ii) when $H(t)$ and $\mathcal{L} (t)$ are such that $\mathbb{K} (t) = 0_M$, 
$\forall t \in [0,\tau]$, leaving a non-trivial holonomic part only; (iii) when $H(t)$ and $\mathcal{L} (t)$ 
are such that $\mathbb{A} (t) = -i A_j (t) \mathbb{P}_j (0)$ and $\mathbb{K} (t) = 
-i K_j (t) \mathbb{P}_j (0)$, $\mathbb{P}_j (0)$ being mutually orthogonal projection matrices 
and $A_j (t),K_j(t)$, $j=1,\ldots,M$, are real-valued functions that are simultaneously 
non-zero for at least part of the time interval $[0,\tau]$, and $A_{j'}(t) \neq A_{j\neq j'}(t)$ 
[$K_{k'} (t) \neq K_{k\neq k'} (t)$] for at least one $j'$ ($k'$). Cases (i) and (ii) have been utilized, e.g., 
in Ref.~\cite{zheng16} and in Refs.~\cite{sjoqvist12,xu15,sjoqvist16,herterich16}, respectively, 
and case (iii) is reminiscent of the idea of unconventional geometric quantum computation 
\cite{zhu03}, which has been considered in the non-Abelian context in Ref.~\cite{zhao23}. 
The typical case, which we label (iv), is when the holonomic and dynamical parts do not separate. 
In Sec.~\ref{sec:IV}, (i)-(iv) are illustrated in a three-level system.

Whether a given solution $\mathbb{W}(\tau)$ separates or not is a gauge invariant property 
of the time evolution. To prove this assertion, consider a change of $M$-frame 
$\mathcal{L}(t) \mapsto \bar{\mathcal{L}}(t)$ under which  
\begin{eqnarray}
\ket{\varphi_k (t)} \mapsto \ket{\bar{\varphi}_k (t)} = \ket{\varphi_j (t)} \mathbb{R}_{jk} (t)
\label{eq:gaugetransformation} 
\end{eqnarray}
for some unitary $M\times M$ matrix $\mathbb{R}(t)$ smoothly dependent on time, 
$\mathbb{W} (\tau) \mapsto \bar{\mathbb{W}} (\tau)$, and $\mathbb{O} (0,\tau) \mapsto 
\bar{\mathbb{O}} (0,\tau)$, so that   
\begin{eqnarray}
\mathbb{U}(\tau,0) & = & \mathbb{O}(0,\tau) \mathbb{W}(\tau) 
\nonumber \\ 
 & \mapsto & \bar{\mathbb{U}}(\tau,0) = 
\bar{\mathbb{O}} (0,\tau) \bar{\mathbb{W}} (\tau) . 
\label{eq:utransform1}
\end{eqnarray}
Now, note that 
\begin{eqnarray}
\bar{\mathbb{U}} (\tau,0) = \mathbb{R}^{\dagger} (0) 
\mathbb{U}(\tau,0) \mathbb{R} (0), 
\label{eq:utransform2}
\end{eqnarray}
since $\mathcal{S}(0)$ is required to coincide with $\mathcal{L}(0)$. We further 
have from Eq.~\eqref{eq:gaugetransformation} that 
$\bar{\mathbb{O}} (0,\tau) = \mathbb{R}^{\dagger} (0) \mathbb{O} (0,\tau)\mathbb{R} (\tau) >0$. 
Together with Eq.~\eqref{eq:inphase}, this implies $\mathbb{R} (\tau) = \mathbb{R} (0)$ and thus  
\begin{eqnarray}
\bar{\mathbb{O}} (0,\tau) = \mathbb{R}^{\dagger} (0) \mathbb{O} (0,\tau)\mathbb{R} (0) . 
\label{eq:otransform}
\end{eqnarray}
By combining Eqs.~\eqref{eq:utransform2} and \eqref{eq:otransform} with 
Eq.~\eqref{eq:utransform1}, we find 
\begin{eqnarray}
\mathbb{R}^{\dagger} (0) 
\mathbb{U}(\tau,0) \mathbb{R} (0) = \mathbb{R}^{\dagger} (0) 
\mathbb{O}(0,\tau) \mathbb{R} (0) \bar{\mathbb{W}} (\tau) . 
\label{eq:intermediate}
\end{eqnarray}
Since $\mathbb{O} (0,\tau)$ is assumed to be strictly positive, it follows that 
$\mathbb{O}^{-1} (0,\tau)$ exists. Thus, $\left[ \mathbb{R}^{\dagger} (0) 
\mathbb{O}(0,\tau) \mathbb{R} (0) \right]^{-1} = \mathbb{R}^{\dagger} (0) 
\mathbb{O}^{-1} (0,\tau) \mathbb{R} (0)$ is well defined and can be used to 
invert Eq.~\eqref{eq:intermediate}, yielding   
\begin{eqnarray}
\bar{\mathbb{W}} (\tau)
 & = & \mathbb{R}^{\dagger} (0) \mathbb{O}^{-1} (0,\tau) \mathbb{R} (0) \mathbb{R}^{\dagger} (0) 
\mathbb{O}(0,\tau) \mathbb{W} (\tau) \mathbb{R} (0) 
\nonumber \\ 
 & = & \mathbb{R}^{\dagger} (0) \mathbb{W} (\tau) 
\mathbb{R} (0) .   
\end{eqnarray}
In the case where $\mathbb{W} (\tau)$ separates,  
we therefore have  
\begin{eqnarray}
 & & 
\mathcal{T}e^{\int_{0}^{\tau} \mathbb{A}(t) dt}  \mathcal{T} 
e^{\int_{0}^{\tau} \mathbb{K}(t) dt} 
\nonumber \\ 
 & \mapsto & \mathbb{R}^{\dagger} (0) 
\mathcal{T}e^{\int_{0}^{\tau} \mathbb{A}(t) dt} \mathbb{R} (0) 
\mathbb{R}^{\dagger} (0)\mathcal{T}e^{\int_{0}^{\tau} 
\mathbb{K}(t) dt} \mathbb{R} (0) ,   
\label{eq:giprod}
\end{eqnarray}
which is a product of the holonomic part $\mathbb{R}^{\dagger} (0) 
\mathcal{T}e^{\int_{0}^{\tau} \mathbb{A}(t) dt} \mathbb{R} (0)$ and dynamical part 
$\mathbb{R}^{\dagger} (0)\mathcal{T}e^{\int_{0}^{\tau} 
\mathbb{K}(t) dt} \mathbb{R} (0)$ in the transformed frame $\bar{\mathcal{L}} (t)$. 
This shows that if $\mathbb{W} (\tau)$ separates in one gauge, it will do so in all gauges. 
In particular, it follows that if $\mathbb{W}(\tau)$ does not separate in a given gauge, it will 
not separate in any gauge. Consequently, (i)-(iii) above constitute the only special cases in 
which separation occurs.

Note that the eigenvalues of $\mathbb{R}^{\dagger} (0) 
\mathcal{T}e^{\int_{0}^{\tau} \mathbb{A}(t) dt} \mathbb{R} (0)$ coincide with those of 
$\mathcal{T}e^{\int_{0}^{\tau} \mathbb{A}(t) dt}$, meaning that the eigenvalues of the 
holonomic part are gauge invariant. Since the eigenvalues can be obtained by 
reconstructing the unitary time evolution matrix using quantum state tomography 
\cite{abdumalikov13,feng13,arroyo14,zu14,zhou17}, the spectrum of the first factor 
in Eq.~\eqref{eq:giprod} is the experimental signature of the holonomy associated 
with the loop $C$ in all gauges.

It is instructive to consider the separation issue for cyclic evolution. In this case, there 
exists a set of vectors $\mathcal{S}(0) = \mathcal{L}(0) = \{ \ket{\mu_j} \}_{j=1}^M$ 
spanning $\mathscr{V}_M (\tau) = \mathscr{V}_M (0)$ and satisfying [repeated 
indices are not summed in Eqs.~\eqref{eq:Ueigen}-\eqref{eq:Acyc}]
\begin{eqnarray}
U(\tau,0) \ket{\mu_j} = e^{i\chi_j} \ket{\mu_j}, 
\label{eq:Ueigen}
\end{eqnarray}
i.e., $\ket{\mu_j}$ and $e^{i\chi_j}$ are eigenvectors and eigenvalues of $U(\tau,0)$. 
We can construct $\mathcal{L}(t)$ as 
\begin{eqnarray}
\ket{\varphi_j (t)} = e^{-if_j (t)} U(t,0) \ket{\mu_j}, 
\label{eq:cyclicsection}
\end{eqnarray}
with real-valued functions $f_j(t)$ such that $f_j (\tau) = \chi_j$ and $f_j (0) = 0$ \cite{remark5}. 
This guarantees the cyclic condition as the accumulated phases precisely cancel the eigenphases 
of $U(\tau,0)$, yielding $\ket{\varphi_j (\tau)} = \ket{\varphi_j (0)} = \ket{\mu_j}$ 
and hence $\mathbb{O} (0,\tau) = \mathbb{I}_M$.
Equation \eqref{eq:cyclicsection} allows us to compute $\mathbb{K}(t)$ and $\mathbb{A}(t)$ in 
the cyclic case as  
\begin{eqnarray}
\mathbb{K}_{jk} (t) = -i \bra{\mu_j} U^{\dagger} (t,0) H(t) U(t,0) \ket{\mu_k} e^{i[f_j(t) - f_k(t)]} , 
 \nonumber \\ 
\label{eq:Kcyc} 
\end{eqnarray}
and 
\begin{eqnarray}
 & & \mathbb{A}_{jk} (t) = i\dot{f}_{j} (t) 
\delta_{jk} e^{i[f_j(t) - f_k(t)]} 
\nonumber \\  
  & + & i \bra{\mu_j} U^{\dagger} (t,0)  H(t) U(t,0) \ket{\mu_k} e^{i[f_j(t) - f_k(t)]} ,   
\label{eq:Acyc}
\end{eqnarray}
where we have used Eq.~\eqref{eq:aeaf} and that $\dot{U}^{\dagger}(t,0) = i U^{\dagger}(t,0) H(t)$.
By comparing Eqs.~\eqref{eq:Kcyc} and \eqref{eq:Acyc}, we find  
\begin{eqnarray}
\mathbb{A} (t) & = & i \, {\rm diag} \{ \dot{f}_1 (t),\ldots, \dot{f}_M (t) \} - \mathbb{K} (t)
\nonumber \\ 
 & \equiv &\mathbb{J}(t) - \mathbb{K} (t) ,
 \label{eq:AD}
\end{eqnarray}
which, combined with Eq.~\eqref{eq:AKcomm}, implies 
the criterion 
\begin{eqnarray}
[\mathbb{J}(t) ,\mathbb{K} (t')] = [\mathbb{K} (t),\mathbb{K} (t')] , \quad \forall t,t' \in [0,\tau], 
\label{eq:criterion}
\end{eqnarray}
for when the holonomic and dynamical parts separate. 
The typical case (iv) 
occurs when $[\mathbb{J}(t) ,\mathbb{K} (t')] \neq [\mathbb{K} (t),\mathbb{K} (t')]$ for 
at least some $t,t' \in [0,\tau]$. We recognize (i)-(iii) as the following three special cases 
where Eq.~\eqref{eq:criterion} holds: (i) $\mathbb{J}(t) = \mathbb{K} (t)$, $\forall t\in[0,\tau]$, 
which implies $\mathbb{A}(t) = 0_M$ by Eq.~\eqref{eq:AD}; (ii) $\mathbb{K} (t) = 0_M$, 
$\forall t\in[0,\tau]$; (iii) $\mathbb{A}(t) = -i \, {\rm diag} \{ A_1 (t), \ldots ,A_M (t) \} \equiv 
\mathbb{A}_d (t)$ and $\mathbb{K} (t) = -i \, {\rm diag} \{ K_1 (t), \ldots ,K_M (t) \} \equiv 
\mathbb{K}_d (t)$, which implies $\mathbb{J} (t) = \mathbb{A}_d (t) + \mathbb{K}_d(t)$. 
Further, note that 
\begin{eqnarray}
\mathbb{U} (\tau,0) = e^{\int_0^{\tau} \mathbb{J} (t) dt} = {\rm diag} \{ e^{i\chi_1}, \ldots, e^{i\chi_M} \}
\end{eqnarray}
in all cases (i)-(iv). We thus find $\mathbb{U} (\tau,0) = e^{\int_0^{\tau} \mathbb{K} (t) dt}$ 
in case (i), $\mathbb{U} (\tau,0) = e^{\int_0^{\tau} \mathbb{A} (t) dt}$ in case (ii), and 
$\mathbb{U} (\tau,0) = e^{\int_0^{\tau} \mathbb{A}_d (t) dt} 
e^{\int_0^{\tau} \mathbb{K}_d (t) dt} = e^{\int_0^{\tau} \mathbb{K}_d (t) dt} 
e^{\int_0^{\tau} \mathbb{A}_d (t) dt}$ in case (iii). Note that no time ordering is needed 
in any of these expressions for $\mathbb{U}(\tau,0)$.   

\begin{figure*}[ht]
\centering
\includegraphics[width=0.6\textwidth]{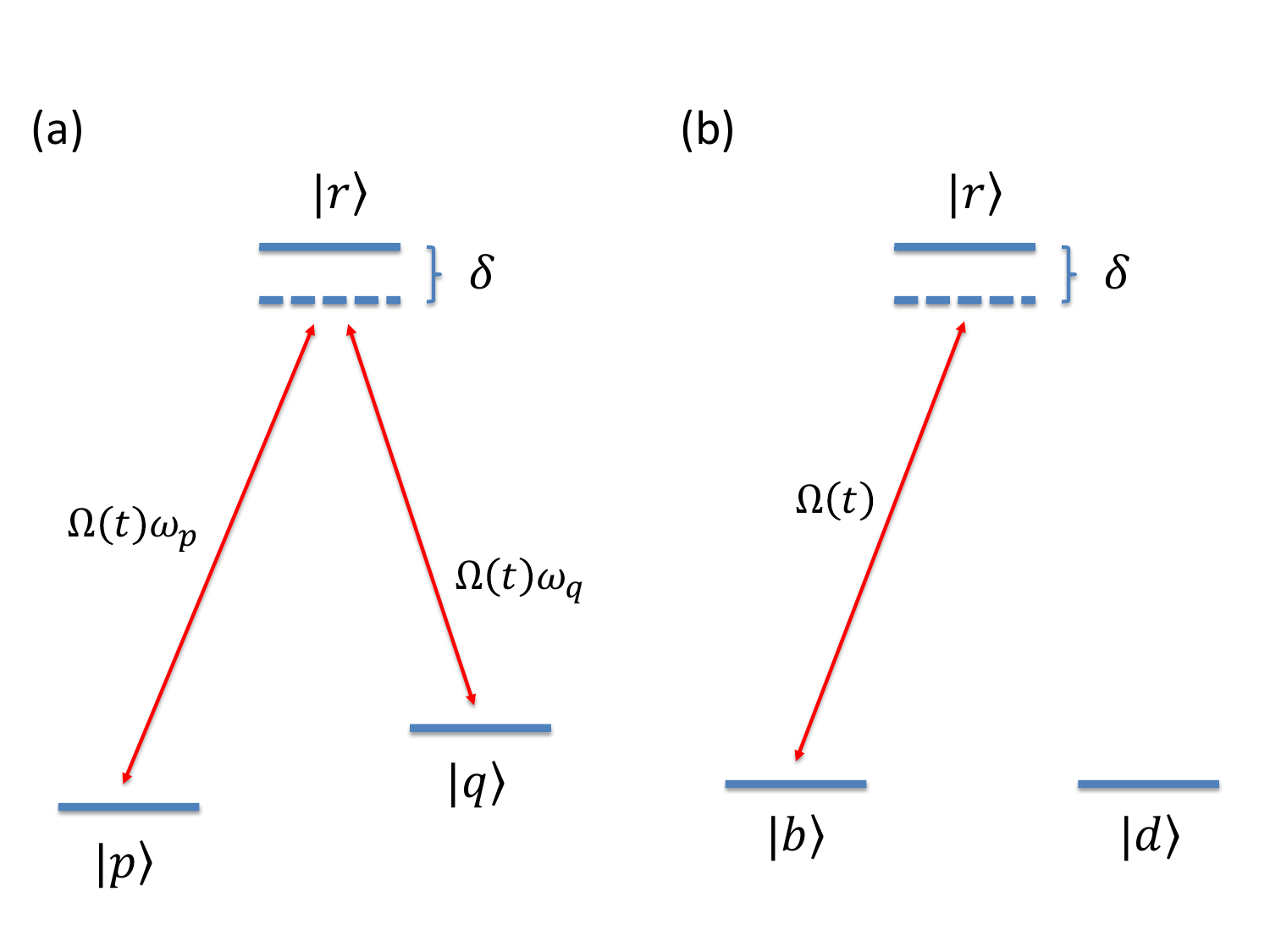}
\caption{(a) Three-level $\Lambda$ configuration with two transitions 
$\ket{p} \leftrightarrow \ket{r}$ and $\ket{q} \leftrightarrow \ket{r}$, $(p,q,r)$ being a 
given permutation of some fixed orthogonal states with labels $(1,2,3)$. The two 
transitions are simultaneously driven by a pair of laser pulses both with real-valued 
envelope $\Omega (t)$, the same detuning $\delta$, and time-independent relative 
amplitude $\omega_p/\omega_q$. (b) The system expressed in the Morris-Shore 
basis \cite{morris83} with $\ket{b} = \omega_p^{\ast} \ket{p} + \omega_q^{\ast} \ket{q}$ 
and $\ket{d} = -\omega_q \ket{p} + \omega_p \ket{q}$.}
\label{fig:1}
\end{figure*}

\section{Physical examples}
\label{sec:IV}
To illustrate cases (i)-(iv) identified 
in Sec.~\ref{sec:IIC}, we consider the `minimal' 
physical setting consisting of a three-level $\Lambda$ configuration, in which two transitions 
$\ket{p} \leftrightarrow \ket{r}$ and $\ket{q} \leftrightarrow \ket{r}$, with $(p,q,r)$ a given 
permutation of some fixed orthogonal states with labels $(1,2,3)$, are simultaneously 
driven by a pair of square-shaped laser pulses both with real-valued envelope 
\begin{eqnarray}
\Omega (t) = \left\{ \begin{array}{ll} 
\Omega_0, & 0 \leq t \leq \tau, \\ 
0, & {\rm otherwise.}
\end{array}
\right.  
\end{eqnarray}
By using the rotating wave approximation in the appropriate rotating frame, the Hamiltonian 
during the pulse becomes time-independent and takes the form  
\begin{eqnarray}
H = \Omega_0 \left( \ket{r} \bra{b} + \ket{b} \bra{r} \right) + 2\delta \ket{r}\bra{r} ,  
\label{eq:ham}
\end{eqnarray}
where $\ket{b} = \omega_p^{\ast} \ket{p} + \omega_q^{\ast} \ket{q}$ is the Morris-Shore 
bright state \cite{morris83}, 
$\omega_p,\omega_q \in \mathbb{C}$ are time-independent laser parameters satisfying 
$|\omega_p|^2 + |\omega_q|^2 = 1$, and we have assumed the two pulses have the 
same detuning $\delta$. The setup is shown in Fig.~\ref{fig:1}. Diagonalization of $H$ yields 
\begin{eqnarray}
H = \mathcal{E}_1 \ket{v_1} \bra{v_1} + \mathcal{E}_2  \ket{v_2} \bra{v_2}  
\end{eqnarray}
with eigenvectors 
\begin{eqnarray}
\ket{v_1} & = & \cos \frac{\gamma}{2} \ket{r} + \sin \frac{\gamma}{2} \ket{b} , 
\nonumber \\ 
\ket{v_2} & = & -\sin \frac{\gamma}{2} \ket{r} + \cos \frac{\gamma}{2} \ket{b} ,  
\end{eqnarray}
and corresponding energy eigenvalues  
\begin{eqnarray}
\mathcal{E}_1 & = & \delta + \sqrt{\delta^2 + \Omega_0^2} \equiv \delta + \dot{\phi}_t, 
\nonumber \\ 
\mathcal{E}_2 & = & \delta - \sqrt{\delta^2 + \Omega_0^2} \equiv \delta - \dot{\phi}_t .
\label{eq:eigenvalues}
\end{eqnarray}
Here, we have introduced $\tan \gamma \equiv \Omega_0/\delta$ and the precession 
angle $\phi_t \equiv \sqrt{\delta^2+\Omega_0^2} \, t$. We note in particular that zero 
detuning corresponds to $\gamma = \pm \frac{\pi}{2}$. The third eigenvector
$\ket{v_0} =  -\omega_q \ket{p} + \omega_p \ket{q}$ with $\mathcal{E}_0 = 0$ is 
the Morris-Shore dark state $\ket{d}$, which is decoupled from the dynamics \cite{morris83}. 

We demonstrate how the special cases (i)-(iii) are obtained in the above described $\Lambda$ system 
by choosing the initial $M$-frame $\mathcal{S} (0) = \mathcal{L} (0)$ in three different ways. 

To illustrate the typical case (iv), we consider a sequence of two $\Lambda$ systems 
with different choices of $(p,q,r)$ so as to correspond to two non-commuting $\Lambda$ 
Hamiltonians. 

\subsection{Case (i)} 
\label{sec:IIIA}
Let $\mathcal{S} (0) = \mathcal{L} (0) = \{ \ket{v_1},\ket{v_2} \}$, which implies 
$\dot{P}_2 (t) = \frac{d}{dt}(\hat{1} - \ket{d}\bra{d}) = 0$, since $\ket{d} \bra{d}$ does 
not evolve under influence of $H$ in Eq.~\eqref{eq:ham}. This allows us to take the 
fixed vectors $\ket{\varphi_1 (t)} = \ket{\varphi_1 (0)} = \ket{v_1}$ and 
$\ket{\varphi_2 (t)} = \ket{\varphi_2 (0)} = \ket{v_2}$ to correspond to the local section 
with overlap matrix $\mathbb{O} (0,\tau) = \mathbb{I}_2$. One therefore obtains 
\begin{eqnarray}
\mathbb{K} (t) = 
-i \begin{pmatrix}
\mathcal{E}_1  & 0 \\      
0 & \mathcal{E}_2      
\end{pmatrix} 
\end{eqnarray}
and $\mathbb{A} (t) = 0_2$, yielding the purely dynamical time evolution matrix \cite{zheng16}
\begin{eqnarray}
\mathbb{U} (\tau,0) = e^{\int_0^{\tau} \mathbb{K} (t) dt} = e^{-i \delta \tau} 
e^{-i\phi_{\tau} \mathbb{Z}} ,
\end{eqnarray} 
where $\mathbb{Z} = 
\begin{pmatrix}
    1 & 0 \\
    0 & -1
\end{pmatrix}$. 

\subsection{Case (ii)} 

\label{sec:IIIB}
Let now $\mathcal{S} (0) = \mathcal{L} (0) = \{ \ket{d}, \ket{b} \}$, which implies 
$\dot{P}_2 (t) \neq 0$, potentially giving rise to a non-trivial holonomy. We choose 
$\mathcal{L} (t) = \{ \ket{\varphi_1 (t)} = \ket{\varphi_1 (0)} = \ket{d} , \ket{\varphi_2 (t)} = 
e^{-i \arg \bra{b} e^{-iHt} \ket{b}} e^{-i Ht} \ket{b}\}$ to correspond to the local section 
with overlap matrix 
\begin{eqnarray}
\mathbb{O} (0,t) = \begin{pmatrix}
1 & 0 \\ 
0 & \sqrt{1 - \sin^2 \gamma \sin^2 \phi_t}
\end{pmatrix} \geq 0. 
\end{eqnarray}
Note that $\mathbb{O} (0,t) > 0$, $\forall t \in [0,\tau]$, if $\sin \gamma \neq \pm 1$, which 
corresponds to non-zero detuning. 

From $\ket{\varphi_{1}(t)} = \ket{d}$ it follows that $\mathbb{K}_{11}(t) = \mathbb{K}_{12}(t) = 
\mathbb{K}_{21}(t) = 0$, and $\ket{\varphi_2 (t)}$ implies $\mathbb{K}_{22}(t) = 
-i \bra{\varphi_2 (t)} H \ket{\varphi_2 (t)} = 
-i \bra{b} e^{iHt} H e^{-iHt} \ket{b} = -i \bra{b} H \ket{b} = 0$. We further have 
\begin{eqnarray}
\mathbb{A} (t) = 
i\begin{pmatrix}
0 & 0 \\ 
0 & \frac{d}{dt} \arg \bra{b} e^{-iHt} \ket{b} 
\end{pmatrix} ,  
\end{eqnarray}
yielding, together with $\mathbb{K}(t) = 0_{2}$, the solution of AE:
\begin{eqnarray}
\mathbb{W} (\tau) & = & e^{\int_0^{\tau} \mathbb{A} (t) dt} 
=  
\begin{pmatrix}
1 & 0 \\ 
0 & e^{i \arg \bra{b} e^{-iH\tau} \ket{b}} 
\end{pmatrix} , 
\end{eqnarray}
where we have used that time ordering is redundant, since  $[\mathbb{A} (t),\mathbb{A} (t')] = 0_2$, 
$\forall t,t' \in [0,\tau]$. Thus, the time evolution matrix becomes 
\begin{eqnarray}
& & \mathbb{U} (\tau,0) = \mathbb{O} (0,\tau) \mathbb{W} (\tau) 
\nonumber \\ 
 & = & 
\begin{pmatrix}
1 & 0 \\ 
0 & \sqrt{1 - \sin^2 \gamma \sin^2 \phi_{\tau}} \, e^{i \arg \bra{b} 
e^{-iH\tau} \ket{b}}
\end{pmatrix} ,  
\nonumber \\  
\end{eqnarray}
where $\arg \bra{b} e^{-iH\tau} \ket{b} = -\delta \tau + \arg \left( \cos \phi_\tau + 
i \cos \gamma \sin \phi_\tau \right)$. Cyclic evolution ($\mathbb{O} (0,\tau) = \mathbb{I}_2$) 
corresponds to $\phi_{\tau} = \pi$ and $\delta\tau = \pi \delta /\sqrt{\delta^2 + \Omega_0^2} = 
\pi \cos \gamma$, which provides a 
realization of the non-trivial holonomy \cite{xu15,sjoqvist16}
\begin{eqnarray}
\mathbb{U} (\tau,0) = \mathbb{W} (\tau) = \begin{pmatrix}
1 & 0 \\ 
0 & e^{-i \pi \left( 1 + \cos \gamma \right)}
\label{eq:Ucycii}
\end{pmatrix} . 
\end{eqnarray} 
Note that $\mathbb{U} (\tau,0)$ is diagonal in the dark–bright basis but, in general, off-diagonal 
in the fixed basis $\{ \ket{p},\ket{q} \}$. This displays the non-Abelian nature of the holonomy: 
two non-commuting time evolution matrices can be realized by applying two consecutive 
pulse-pairs with different choices of laser parameters $\omega_p,\omega_q$ \cite{sjoqvist12}. 

\subsection{Case (iii)}  
Here, we make use of the energy eigenvectors by choosing $\mathcal{S} (0) = 
\mathcal{L} (0) = \{ \ket{v_0} = \ket{d}, \cos \frac{\xi}{2} \ket{v_1} + 
\sin \frac{\xi}{2} \ket{v_2} \}$, $\xi \in [0,\pi]$, which implies $\dot{P}_2 (t) \neq 0$ 
so as to allow for a non-trivial holonomy. The choice $\mathcal{L} (t) = 
\{ \ket{\varphi_1 (t)} = \ket{\varphi_1 (0)} = 
\ket{d} , \ket{\varphi_2 (t)} =  e^{-i \arg \bra{\psi_2 (0)} \psi_2 (t) \rangle} \ket{\psi_2 (t)} \}$ 
with $\ket{\psi_2 (t)} = e^{-iHt} \ket{\psi_2 (0)} = e^{-i \mathcal{E}_1 t} \cos \frac{\xi}{2} \ket{v_1} + 
e^{-i \mathcal{E}_2 t} \sin \frac{\xi}{2} \ket{v_2}$, corresponds to a local section, since 
\begin{eqnarray}
\mathbb{O} (0,t) = \begin{pmatrix} 1 & 0 \\ 
0 & \sqrt{1-\sin^2 \xi \sin^2 \phi_t}  
\end{pmatrix} \geq 0.
\label{eq:Oiii}
\end{eqnarray}
The inequality becomes strict for $\xi \neq \pm \frac{\pi}{2}$. 

We can now calculate $\mathbb{A} (t)$ and $\mathbb{K} (t)$ by using $\mathcal{L} (t)$. 
One finds: 
\begin{eqnarray}
\mathbb{A} (t) 
= i \begin{pmatrix} 
0 & 0 \\ 
0 & \frac{d}{dt} \arg \langle \psi_2 (0) \ket{\psi_2 (t)} + \delta + \dot{\phi}_t \cos \xi 
\end{pmatrix} 
\label{eq:Aiii}
\end{eqnarray}
and  
\begin{eqnarray}
\mathbb{K} (t) = 
-i \begin{pmatrix} 
0 & 0 \\ 
0 & \delta + \dot{\phi}_t \cos \xi 
\end{pmatrix} .  
\label{eq:Fiii}
\end{eqnarray}
Conceptually, this corresponds to the idea of unconventional geometric quantum computation \cite{zhu03} 
in the holonomic setting \cite{zhao23}, in which $\mathbb{K} (t) = -g(t) \mathbb{A}(t)$, here with 
\begin{eqnarray}
g(t) = \frac{\delta + \dot{\phi}_t \cos \xi}{\frac{d}{dt} \arg \langle \psi_2 (0) 
\ket{\psi_2 (t)} + \delta + \dot{\phi}_t \cos \xi} .
\end{eqnarray}
Equations \eqref{eq:Aiii} and \eqref{eq:Fiii} show that $[\mathbb{A} (t), \mathbb{K} (t') ] = 0_2$, 
$\forall t,t' \in [0,\tau]$. Thus, the solution of AE separates: 
\begin{eqnarray}
\mathbb{W} (\tau) & = & e^{\int_0^{\tau} \mathbb{A} (t) dt} e^{\int_0^{\tau} 
\mathbb{K} (t) dt} 
\nonumber \\ 
 & = & \begin{pmatrix} 1 & 0 \\ 
0 & e^{i \left[ \arg(\cos\phi_\tau - i\cos\xi\sin\phi_\tau) + \phi_{\tau} \cos \xi \right]}
\end{pmatrix}
\nonumber \\ 
 & & \times 
\begin{pmatrix} 1 & 0 \\ 
0 & e^{-i \left( \delta \tau + \phi_{\tau} \cos \xi \right)} 
\end{pmatrix} , 
\label{eq:wIIIC}
\end{eqnarray}
where time ordering is not needed, since $[\mathbb{A} (t),\mathbb{A} (t')] = 
[\mathbb{K} (t),\mathbb{K} (t')] = 0_{2}$, $\forall t,t' \in [0,\tau]$, and we have used that 
$\arg \langle \psi_2 (0) \ket{\psi_2 (\tau)} = -\delta\tau + \arg(\cos\phi_\tau - i\cos\xi\sin\phi_\tau)$. 
 
Note that we recover the example given in Sec.~\ref{sec:IIIB} when $\xi = \pi - \gamma$.
This can be seen by noting that 
\begin{eqnarray} 
\ket{\psi_2 (0)} & = & 
\cos \left( \frac{\pi - \gamma}{2} \right) \ket{v_1} + \sin\left( \frac{\pi - \gamma}{2} \right) \ket{v_2} 
\nonumber \\ 
 & = & \sin \frac{\gamma}{2} \ket{v_1} + 
\cos \frac{\gamma}{2} \ket{v_2} = \ket{b} 
\end{eqnarray}
and 
\begin{eqnarray}
 \delta + \dot{\phi}_t \cos (\pi - \gamma) = 
\delta - \dot{\phi}_t \cos \gamma = 0  , 
\end{eqnarray}
yielding $\mathbb{A}_{22} (t) = i\frac{d}{dt}\arg \bra{b} e^{-iHt} \ket{b}$ and 
$\mathbb{K}_{22} (t) = 0$ in Eqs.~\eqref{eq:Aiii} and \eqref{eq:Fiii}, respectively. 
Furthermore, note that $\xi = 0$ ($\xi = \pi$) correspond to $\mathcal{S} (0) = 
\mathcal{L} (0) = \{ \ket{d},\ket{v_1} \}$ ($\mathcal{S} (0) = \mathcal{L} (0) = 
\{ \ket{d},\ket{v_2} \}$). Just as in the example given in Sec.~\ref{sec:IIIA}, these 
two initial frames give rise to purely dynamical time evolution. In this sense, the 
setups discussed in Secs.~\ref{sec:IIIA} and \ref{sec:IIIB} are special cases of the 
setting discussed here, obtained by selecting the value of $\xi$ appropriately.

It is instructive to consider the cyclic case, in which $\mathbb{U} (\tau,0)$ coincides 
$\mathbb{W} (\tau)$. As is evident from Eq.~\eqref{eq:Oiii}, cyclic evolution corresponds 
to $\phi_{\tau} = \pi$ (the already mentioned cases $\xi = 0,\pi$ also trivially corresponds 
to cyclic evolution, since $\mathbb{O}(0,t) = \mathbb{I}_{2}$, $\forall t\in[0,\tau]$), which 
implies $\delta \tau = \pi \cos \gamma$. Upon insertion into Eq.~\eqref{eq:wIIIC}, this 
results in the time evolution matrix \begin{eqnarray}
\mathbb{U} (\tau,0) 
 & = & \begin{pmatrix} 1 & 0 \\ 
0 & e^{-i \pi (1 - \cos \xi)}
\end{pmatrix}
\nonumber \\ 
 & & \times 
\begin{pmatrix} 1 & 0 \\ 
0 & e^{-i \pi (\cos \gamma + \cos \xi)} 
\end{pmatrix} ,   
\label{eq:productformC}
\end{eqnarray}
which again is consistent with the example given in Sec.~\ref{sec:IIIB}, as it reduces 
to Eq.~\eqref{eq:Ucycii} for the special choice $\xi = \pi - \gamma$. Conversely, for a 
given $\xi$, the product form displayed in Eq.~\eqref{eq:productformC} makes it 
possible to realize experimentally a pure holonomy 
\begin{eqnarray}
\mathbb{U} (\tau,0) = \begin{pmatrix} 1 & 0 \\ 
0 & e^{-i \pi (1 - \cos \xi)}
\end{pmatrix}
\end{eqnarray} 
by independently tuning $\gamma$ until it coincides with $\pi - \xi$.

\begin{figure*}[ht]
\centering
\includegraphics[width=0.6\textwidth]{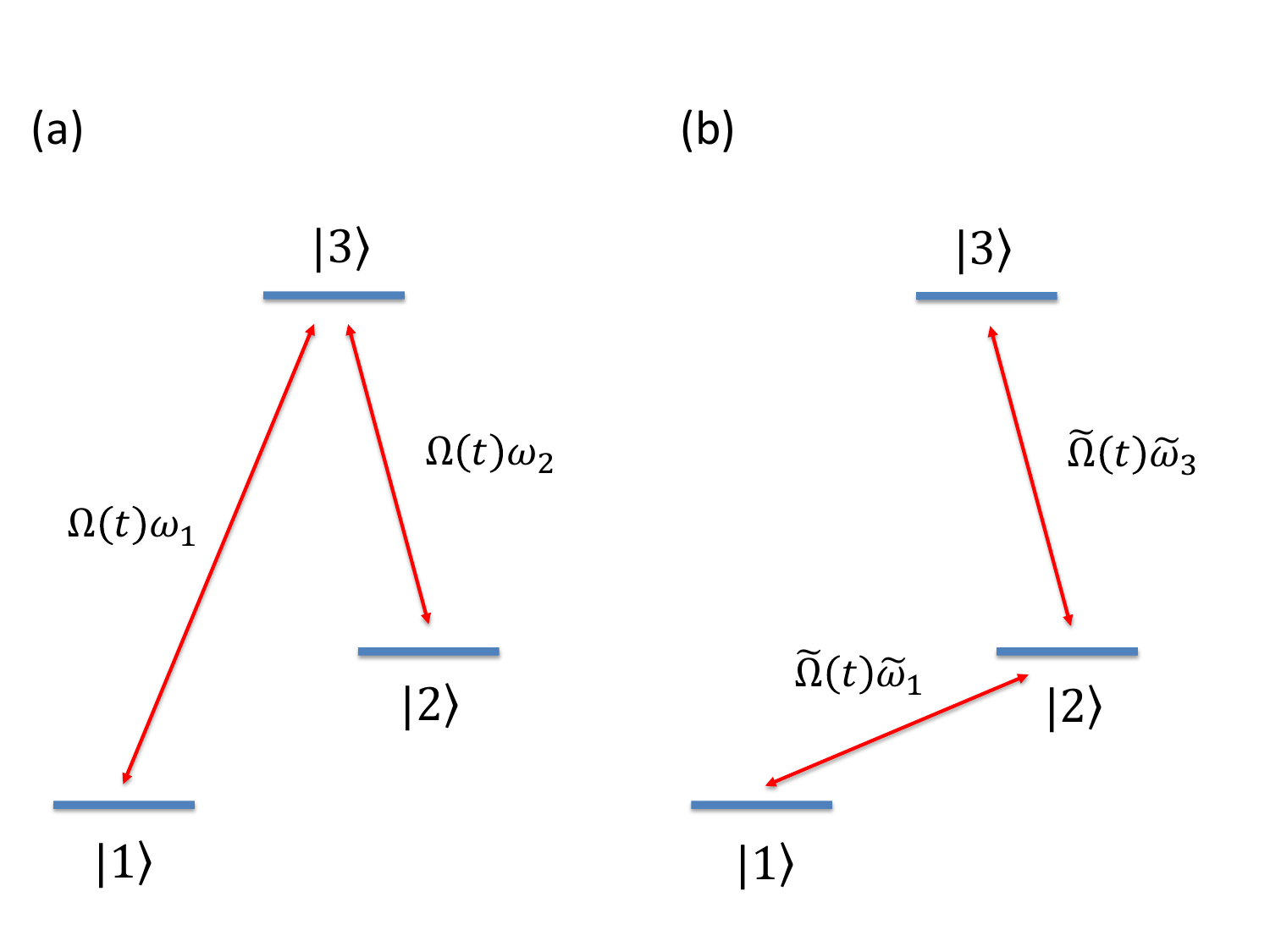}
\caption{Two different resonant ($\delta = 0$) $\Lambda$-type coupling structures 
applied sequentially: (a) a pair of laser pulses simultaneously driving the transitions 
$\ket{1} \leftrightarrow \ket{3}$ and $\ket{2} \leftrightarrow \ket{3}$ for $t \in [0,t_1]$, 
followed by (b) a pair of laser pulses simultaneously driving the transitions 
$\ket{1} \leftrightarrow \ket{2}$ and $\ket{3} \leftrightarrow \ket{2}$ for $t \in [t_1,\tau]$. 
The resulting holonomic and dynamical parts do not separate on $[0,\tau]$ due to 
the non-commuting nature of the Hamiltonians associated with the coupling structures.}
\label{fig:2}
\end{figure*}

\subsection{Case (iv)}  
To illustrate the typical case where the holonomic and dynamical parts do not separate, 
we consider two resonant ($\delta = 0$) $\Lambda$-type Hamiltonians $H(t) = \Omega (t) H_0$ 
and $\tilde{H}(t) = \tilde{\Omega} (t) \tilde{H}_0$ such that $[H_0,\tilde{H}_0] \neq 0$. 
To achieve non-commutativity, we associate $H_0$ with a coupling structure consisting 
of the transitions $\{ \ket{p},\ket{q} \} \leftrightarrow \ket{r}$, and $\tilde{H}_0$ similarly 
with $\{ \ket{\tilde{p}},\ket{\tilde{q}} \} \leftrightarrow \ket{\tilde{r}}$, $(p,q,r)$ and 
$(\tilde{p},\tilde{q},\tilde{r})$ being different permutations of $(1,2,3)$. Note that, by 
using resonant pulses, any shape of the envelope functions $\Omega (t)$ and 
$\tilde{\Omega} (t)$ can be used. We focus on the cyclic case. 

For notational simplicity, we make the specific choices $(p,q,r)=(1,2,3)$ and 
$(\tilde{p},\tilde{q},\tilde{r})=(1,3,2)$, i.e., 
\begin{eqnarray}
H_0 & = & \ket{3} \bra{b} + \ket{b} \bra{3} , 
\nonumber \\ 
\tilde{H}_0 & = & \ket{2} \bra{\tilde{b}} + \ket{\tilde{b}} \bra{2}   
\end{eqnarray}  
with $\ket{b} = \omega_1^{\ast} \ket{1} + \omega_2^{\ast} \ket{2}$ and $\ket{\tilde{b}} = 
\tilde{\omega}_1^{\ast} \ket{1} + \tilde{\omega}_3^{\ast} \ket{3}$, applied sequentially 
on time intervals $[0,t_1]$ and $[t_1,\tau]$, respectively. The setup is illustrated in 
Fig~\ref{fig:2}. By assuming $\pi$-pulses ($\int_0^{t_1} \Omega (t) dt = \int_{t_1}^{\tau} 
\tilde{\Omega} (t) dt = \pi$), we obtain the combined unitary operator \cite{alves22} 
\begin{eqnarray}
& & \tilde{V}(\tau,t_1) V(t_1,0) = e^{-i\pi \tilde{H}_0} e^{-i\pi H_0} 
\nonumber \\ 
 & = & 
\big( -\ket{2} \bra{2} + \ket{\tilde{d}} \bra{\tilde{d}} - 
\ket{\tilde{b}} \bra{\tilde{b}} \big)
\nonumber \\ 
 & & \times  
\big( -\ket{3} \bra{3} + \ket{d} \bra{d} - \ket{b} \bra{b} \big)   
\label{eq:combuni} 
\end{eqnarray}
with $\ket{d} = -\omega_2 \ket{1} + \omega_1 \ket{2}$ and $\ket{\tilde{d}} = 
-\tilde{\omega}_3 \ket{1} + \tilde{\omega}_1 \ket{3}$. Here, one can consider various 
combinations of laser parameters $\{ \omega_1, \omega_2, \tilde{\omega}_1, \tilde{\omega}_3\}$ 
and find all sorts of different `subspace unitaries' by factoring out one the eigenvectors 
of the combined unitary operator, say $\ket{\mu_3}$,
so as to implement cyclic evolution $\mathscr{V}_2 (\tau) = \mathscr{V}_2 (0) = 
{\rm Span} \{ \ket{\mu_1},\ket{\mu_2} \}$. The resulting time evolution operator 
restricted to the subspace $\mathscr{V}_2 (0)$ is the partial isometry 
\begin{eqnarray}
U (\tau,0) & = & \tilde{V} (\tau,t_1) V(t_1,0)  \left( \hat{1}-\ket{\mu_3}\bra{\mu_3} \right)
\nonumber \\ 
 & = & e^{i\chi_1} \ket{\mu_1} \bra{\mu_1} + e^{i\chi_2} \ket{\mu_2} \bra{\mu_2}  . 
\end{eqnarray}  
To analyze the issue of separation by purely analytical means in this generic case is 
however surprisingly difficult. 
Therefore, we focus on one very specific parameter choice that yields a non-trivial time
evolution  (i.e., $[H_0,\tilde{H}_0] \neq 0$), for which $\mathbb{A}(t)$ and $\mathbb{K}(t)$ 
can be calculated analytically and shown to be non-commuting. 

To this end, we take $\omega_1 = -\omega_2 = \tilde{\omega}_1 = -\tilde{\omega}_3 = 
\frac{1}{\sqrt{2}}$, which yields  
\begin{eqnarray}
\tilde{V} (\tau,t_1) V (t_1,0) =  
- \ket{1}\bra{3} + \ket{3}\bra{2} - \ket{2}\bra{1}  
\end{eqnarray}
with eigenvalues $e^{i\chi_j} = \left( e^{i2\pi/3} \right)^{j-1} \equiv \eta^{j-1}$, and corresponding 
eigenvectors 
\begin{eqnarray}
\ket{\mu_j} = \frac{1}{\sqrt{3}} 
\left(-\eta^{j-1} \ket{1} + \ket{2} +  \eta^{2(j-1)} \ket{3}\right),     
\end{eqnarray}
$j=1,2,3$. We choose $\mathcal{S}(0) = \mathcal{L}(0) =  \{ \ket{\mu_1},\ket{\mu_2} \}$, i.e., 
\begin{eqnarray}
U (\tau,0) = \ket{\mu_1} \bra{\mu_1} + e^{i2\pi/3} \ket{\mu_2} \bra{\mu_2} 
\end{eqnarray} 
with $\ket{\mu_1} = \frac{1}{\sqrt{3}} \left(- \ket{1} + \ket{2} + \ket{3}\right)$ and 
$\ket{\mu_2} = \frac{1}{\sqrt{3}} \left(-\eta \ket{1} + \ket{2} +  \eta^2 \ket{3}\right)$.

One can note that $\mathscr{V}_2 (0) = {\rm Span} \{ \ket{\mu_1},\ket{\mu_2} \}$ 
does not coincide with the subspaces ${\rm Span} \{ \ket{1},\ket{2} \}$ and  
${\rm Span} \{ \ket{1},\ket{3} \}$. The evolution of these two latter subspaces, governed 
by $V (t,0)$ and $\tilde{V} (t,t_1)$, respectively, are associated with a trivial dynamical part, 
and thereby pure holonomy \cite{sjoqvist12}. Thus, the evolution 
$\mathscr{V}_2 (0) \rightarrow \mathscr{V}_2 (t)$ acquires non-trivial holonomic 
and dynamical contributions. These contributions are not expected to separate after 
completing the loop $C$ in the Grassmannian $\mathscr{G} (3;2)$ at $t=\tau$.

To demonstrate this latter point explicitly, we calculate the corresponding $\mathbb{A} (t)$ 
and $\mathbb{K} (t')$ and show that they indeed do not commute for general $t,t' \in [0,\tau]$. 
First, $V (t_1,0)$ is generated by the Hamiltonian 
\begin{eqnarray}
H(t) & = & \Omega (t) \frac{1}{\sqrt{2}} \left( \ket{3} \bra{1} - \ket{3} \bra{2} + {\rm H.c.} \right) 
\nonumber \\ 
 & \equiv & \Omega (t) H_0,  
\end{eqnarray}
and $\tilde{V} (\tau,t_1)$ is generated by the Hamiltonian 
\begin{eqnarray}
\tilde{H}(t) & = & \tilde{\Omega} (t) \frac{1}{\sqrt{2}} \left( \ket{2} \bra{1} - \ket{2} \bra{3} + {\rm H.c.} \right)  
\nonumber \\ 
 & \equiv & \tilde{\Omega} (t) \tilde{H}_0,
\end{eqnarray}
with the real-valued pulse envelopes $\Omega (t)$ and $\tilde{\Omega} (t)$ being non-zero 
on $[0,t_1]$ and $[t_1,\tau]$, respectively, such that 
$\int_0^{t_1} \Omega (t) dt = \int_{t_1}^{\tau} \tilde{\Omega} (t) dt = \pi$. We verify that 
$[H_0,\tilde{H}_0] = \frac{1}{2} (-\ket{1} \bra{2} +\ket{1} \bra{3} -\ket{2} \bra{3} - {\rm H.c.}) \neq 0$. We have 
\begin{eqnarray}
V(t,0) & = & e^{-ia_t H_0} 
\nonumber \\ 
 & = & \frac{1}{2} \big( \ket{1} + \ket{2} 
\big) \big( \bra{1} + \bra{2} \big)
\nonumber \\ 
 & & + e^{-ia_t} \ket{v_+} \bra{v_+} 
+ e^{ia_t} \ket{v_-} \bra{v_-} 
\end{eqnarray}
with $\ket{v_{\pm}} = \frac{1}{2} \big( \ket{1} - \ket{2} \big) \pm \frac{1}{\sqrt{2}}\ket{3}$ and $a_t = 
\int_0^{t} \Omega (s) ds$; similarly 
\begin{eqnarray}
\tilde{V}(t,t_1) & = & e^{-i\tilde{a}_t \tilde{H}_0} 
\nonumber \\ 
 & = & \frac{1}{2} \big( \ket{1} + \ket{3} 
\big) \big( \bra{1} + \bra{3} \big)
\nonumber \\ 
 & & + e^{-i\tilde{a}_t} \ket{\tilde{v}_+} \bra{\tilde{v}_+} 
+ e^{i\tilde{a}_t} \ket{\tilde{v}_-} \bra{\tilde{v}_-} 
\end{eqnarray}
with $\ket{\tilde{v}_{\pm}} = \frac{1}{2} \big( \ket{1} - \ket{3} \big) \pm \frac{1}{\sqrt{2}} \ket{2}$ 
and $\tilde{a}_t = \int_{t_1}^{t} \tilde{\Omega} (s) ds$. 

The local section $\mathcal{L} (t)$ is chosen as 
\begin{eqnarray}
\ket{\varphi_1 (t)} & = & U (t,0) \ket{\mu_1}, 
\nonumber \\ 
\ket{\varphi_2 (t)} & = & e^{-if(t)} U (t,0) \ket{\mu_2} 
\end{eqnarray}
with $f(t)$ any real-valued function such that $f(\tau) = \frac{2\pi}{3}$ and $f(0) = 0$. Here, 
$U(t,0) = V_{\rm tot} (t,0) (\hat{1} - \ket{\mu_3} \bra{\mu_3})$, where 
\begin{eqnarray}
V_{\rm tot} (t,0) = \left\{ \begin{array}{ll} 
V(t,0), & t\in [0,t_1] , \\ 
\tilde{V}(t,t_1) V(t_1,0), & t \in [t_1,\tau], 
\end{array} \right.  
\end{eqnarray}
is the time evolution operator on the full Hilbert space that coincides with the combined 
unitary $\tilde{V}(\tau,t_1) V(t_1,0)$ in Eq.~\eqref{eq:combuni} at $t=\tau$. Indeed, one 
may verify that $\ket{\varphi_j (\tau)} = \ket{\varphi_j (0)} = \ket{\mu_j}$, $j=1,2$, ensuring 
that $\mathbb{O} (0,\tau) = \mathbb{I}_2$ (cyclic evolution). 

Let us first consider $\mathbb{K}(t)$. We have,  
\begin{widetext}
\begin{eqnarray}
\mathbb{K}_{jk} (t) = \left\{ \begin{array}{ll} -i \bra{\varphi_j (t)} H(t) \ket{\varphi_k (t)} = 
-i \Omega (t) \bra{\mu_j (t)} H_0 \ket{\mu_k (t)}, & t\in [0,t_1], \\ 
-i \bra{\varphi_j (t)} \tilde{H}(t) \ket{\varphi_k (t)} = -i \tilde{\Omega} (t) \bra{\mu_j (t)} 
V^{\dagger} (t_1,0) \tilde{H}_0 
V(t_1,0) \ket{\mu_k (t)} , & t\in [t_1,\tau] , 
\end{array} \right. 
\end{eqnarray}
\end{widetext}
where we have defined $\ket{\mu_1 (t)} \equiv \ket{\mu_1}$ and $\ket{\mu_2 (t)} \equiv 
e^{-if(t)} \ket{\mu_2}$, and 
used that $[H_0,V(t,0)] = [\tilde{H}_0,\tilde{V}(t,t_1)]=0$. By evaluating the matrix elements, we find
\begin{eqnarray}
\mathbb{K} (t) = \left\{ \begin{array}{ll} 
i \frac{\Omega (t)}{3\sqrt{2}} 
\begin{pmatrix}
4 & \eta^2 e^{-if(t)} \\ 
 \eta e^{if(t)} & -2
\end{pmatrix}\!,  & t\in [0,t_1], \\ 
i \frac{\tilde{\Omega} (t)}{3\sqrt{2}} \begin{pmatrix}
4 & \eta e^{-if(t)} \\ 
\eta^2 e^{if(t)} & -2
\end{pmatrix}\!\!, & t\in [t_1,\tau] ,  
\end{array} \right. 
\end{eqnarray}
where we have used that $a_{t_1} = \pi$, and the identity $1+\eta + \eta^2 = 0$. 

Next, we find $\mathbb{A}(t)$ by using $f_1(t)=0$ and $f_2(t)=f(t)$ in Eq.~\eqref{eq:AD}, 
yielding $\mathbb{J} (t) = i \, {\rm diag} \{ 0,\dot{f} (t) \}$, i.e., 
\begin{eqnarray}
\mathbb{A} (t) = i\dot{f} (t) \mathbb{P}_2 - 
\mathbb{K} (t) , 
\label{eq:A}
\end{eqnarray}
where $\mathbb{P}_2 = \begin{pmatrix}
0 & 0 \\ 
0 & 1
\end{pmatrix}$. 
This implies, for arbitrary $t,t' \in [0,\tau]$, that 
\begin{eqnarray}
[\mathbb{A} (t), \mathbb{K} (t')] = i\dot{f} (t)  [\mathbb{P}_2, \mathbb{K} (t')] -[\mathbb{K} (t), \mathbb{K} (t')] 
\end{eqnarray}
with the terms on the right-hand side generally non-vanishing and non-cancelling. 
To illustrate this latter point, it is sufficient to consider $t'=t \in [0,t_1]$ for some $t$ at which 
$\Omega (t) \neq 0$. For this case, we obtain  
\begin{eqnarray}
[\mathbb{A} (t), \mathbb{K} (t)] & = & i\dot{f} (t)  [\mathbb{P}_2, \mathbb{K} (t)] 
\nonumber \\ 
 & = &  
-\frac{\dot{f} (t) \Omega (t)}{3\sqrt{2}} 
\begin{pmatrix}
0 & -\eta^2 e^{-if(t)} \\ 
\eta e^{if(t)} & 0
\end{pmatrix}\! , 
\label{eq:commutAK}
\end{eqnarray}
which, by choosing the function $f(t)$ so that $\dot{f} (t) \neq 0$, does not vanish. We thus 
conclude that the unitaries $\mathcal{T} 
e^{\int_{0}^{\tau} \mathbb{A}(t) dt}$ and $\mathcal{T} 
e^{\int_{0}^{\tau} \mathbb{K}(t) dt}$ do not separate in the time evolution
\begin{eqnarray}
\mathbb{U}(\tau,0) & =  & \mathcal{T} 
e^{\int_{0}^{\tau} [\mathbb{A}(t) + \mathbb{K}(t) ]dt} 
\nonumber \\ 
 & = & e^{i\int_{0}^{\tau} \dot{f} (t) dt \, \mathbb{P}_2} = 
\begin{pmatrix} 
1 & 0 \\ 
0 & e^{i2\pi/3}
\end{pmatrix}\! ,  
\label{eq:last}
\end{eqnarray}
where we have used Eq.~\eqref{eq:A} and $\int_{0}^{\tau} \dot{f} (t) dt = 
f(\tau) - f(0) = \frac{2\pi}{3}$. 

To see the physical significance of the above, recall that the gauge invariant 
eigenvalues of $\mathcal{T} e^{\int_{0}^{\tau} \mathbb{A}(t) dt}$, in the special case 
where the dynamical part can 
independently be made trivial, are found by performing quantum process tomography 
to reconstruct the time evolution matrix $\mathbb{U}(\tau,0)$ experimentally. As a 
consequence of the gauge invariant non-separability displayed in Eq.~\eqref{eq:last}, 
on the other hand, the holonomic part cannot be extracted from $\mathbb{U}(\tau,0)$ 
by independently tuning the dynamical part. 
Thus, the non-adiabatic quantum holonomy in this system setup cannot be measured.

\section{Conclusions}
\label{sec:V}
Separation of non-adiabatic Schr\"odinger-type quantum time evolution of subspaces 
into holonomic and dynamical parts has been studied. We have established that such 
separation only occurs in highly restrictive situations. The special cases in which separation 
does manifest have been identified and found to be essentially covered by existing schemes 
in the literature. We have proved that 
separability is a gauge invariant property of quantum time evolution. 
All qualitatively different special cases have been exemplified within a single $\Lambda$ 
system setup. The typical case, in which the holonomic and dynamical parts do not separate, 
has been illustrated by sequentially applying two non-commuting $\Lambda$-type Hamiltonians. 

While our work completes the analysis of Schr\"odinger evolution of subspaces, the issue of 
separation appears in other contexts involving quantum holonomies. These include, e.g., 
the Uhlmann holonomy \cite{uhlmann86,uhlmann93} of mixed quantum states, and the 
relation between `direct' and `iterative' holonomy associated with discrete sequences of 
subspaces of dimension $M \geq 2$ \cite{sjoqvist06,oi14}. 

\section*{Acknowledgements}
E. S. acknowledges financial support from the Swedish Research Council (VR) through 
Grant No. 2025-05249.


\begin{thebibliography}{99}
\bibitem{remark1} To see this, first note that by differentiating Eq.~\eqref{eq:ytu} with respect 
to time and using Eq.~\eqref{eq:se}, it follows that the time evolution operator satisfies the 
Schr\"odinger equation $\dot{U}(t,0) = -iH(t)U(t,0)$. By utilizing  the identity 
$U(t,0) = P_{M}(t)U(t,0)$ and Eq.~\eqref{eq:vonNeumann} to rewrite the right-hand side, 
we find that $\dot{U}(t,0) = -iH(t)P_{M}(t)U(t,0) = [\dot{P}_{M} (t) - iP_{M}(t)H(t)]U(t,0) = 
[\dot{P}_{M} (t) - iP_{M}(t)H(t)P_{M}(t)]U(t,0)$.
\bibitem{remark2} The more general case $\mathbb{O} (0,\tau) \geq 0$ leads to the concept 
of partial holonomy \cite{kult06}. We avoid this unnecessary complication, by assuming strict 
inequality in Eq.~\eqref{eq:inphase}.
\bibitem{remark3} To see this, we note that \cite{anandan88} 
$\mathcal{T} e^{\int_{0}^{\tau} \mathbb{A}(t) dt} = \mathcal{P} e^{\int_C A}$ with 
$A_{jk} = \langle d\varphi_j \ket{\varphi_k}$ being a matrix-valued connection 1-form 
and $\mathcal{P}$ path ordering along $C$ in $\mathscr{G}(N;M)$.
\bibitem{remark4} Note that we equally well could have used $\mathcal{S}(\tau)$, since 
$\bra{\psi_j (\tau)} U(\tau,0) \ket{\psi_k (\tau)} = \bra{\psi_j (\tau)} \psi_l (\tau) \rangle 
\langle \psi_l (0) \ket{\psi_k (\tau)} = \langle \psi_j (0) \ket{\psi_k (\tau)} = \mathbb{U}_{jk} (\tau,0)$.
\bibitem{remark5} A choice $\bar{f}_j (t)$ with a non-zero $\bar{f}_j (0)$ for a specific 
$j$ can always be transformed into $f_j (t)$ such that $f_j(0) = 0$ by a phase shift: 
$e^{-i\bar{f}_j(t)} U (t,0) \ket{\bar{\mu}_j} = e^{-i[\bar{f}_j(t)-\bar{f}_j(0)]} U(t,0) 
e^{-i\bar{f}_j(0)} \ket{\bar{\mu}_j} \equiv e^{-if_j(t)} U(t,0) \ket{\mu_j}$ with 
$f_j(t) \equiv \bar{f}_j(t) - \bar{f}_j(0)$ and $\ket{\mu_j} \equiv e^{-i\bar{f}_j(0)} \ket{\bar{\mu}_j}$.
\end{thebibliography}
\end{document}